\documentclass[runningheads]{llncs}
\usepackage[T1]{fontenc}
\usepackage{graphicx} 
\usepackage[table,xcdraw]{xcolor} 

\title{Uncut Gem - \\ An Open-Source Hackable Quantum Sensor}

\author{Mark Carney$^\dagger$ \and Victoria Kumaran }
\author{Mark Carney\inst{1}\orcidID{0000-0001-9372-9033} \and
Victoria Kumaran\inst{1}}

\authorrunning{Carney \& Kumaran}

\institute{Quantum Village Inc., Delaware, USA and London, UK 
\email{\{mark,victoria\}@quantumvillage.org} \\ \url{https://quantumvillage.org/}}

\date{April 2025}

\begin{document}

\maketitle

\begin{abstract}
This work presents an overview of our fully open-source, hackable quantum sensor platform based on nitrogen-vacancy (NV) center diamond magnetometry. This initiative aims to democratize access to quantum sensing by providing a comprehensive, modular, and cost-effective system. The design leverages consumer off-the-shelf (COTS) components in a novel hardware configuration, complemented by open-source firmware written in the Arduino IDE, facilitating portability, ease of customization, and future-proofing the design. By lowering the barriers to entry, our sensor serves as a compact platform for education, research, and innovation in quantum technologies, embodying the ethos of open science and community-driven development.

\keywords{Quantum sensing \and Open Source}
\end{abstract}

\section{Introduction}

Quantum technology needs an `Apple II' moment. By this, we refer to the importance that the Apple II holds in the history of the desktop/personal computer. From the first spreadsheet, {\it VisiCalc}, to the first graphical adventure game {\it Mystery House}, the Apple II was a consumer-grade platform that enabled much of the modern technological world we are now so familiar with \cite{Nooney2023}. 

In a similar vein, the release of the Raspberry Pi was viewed as a key enabling technology in the development of the Internet of Things, driving the decentralization of compute in society even further and giving prominence to the Single Board Computer (SBC) \cite{Johnston2017}. Our work has taken inspiration from this concept and applied this reasoning to Quantum Technologies, and sensing in particular. How can we push the boundary of what is conventionally considered to be possible? 

Open Source software has, in the first quarter of the 21st Century, formed the bedrock of much of our technology stacks and infrastructure. Hoffman {\it et al.} \cite{Hoffmann2024} determined that companies would spend around 3.5$\times$ more on software annually if open-source options did not exist. With this, the importance of open sourcing the key parts of the quantum technology stack become apparent - it drives adoption upwards whilst driving costs down \cite{Hoffmann2024}. This sentiment of increasing openness, interoperability, and institutional momentum is echoed by other open source projects and studies in quantum technology, see \cite{Blau2025,Shammah2024}.

Stegemann {\it et al.} \cite{Stegemann2023} managed to get the overall cost of an NV Centre diamond experimentation platform down to under €200, utilising COTS parts and a 3D printer. This made our challenge apparent - could we reduce this cost down even further? And could we fill in some of the gaps and create a full-stack device that is fully open-source, moving away from the optics table and towards a hand-held form factor? 

We first began designing and developing devices for quantum sensing and magnetometry back in 2023, and this project has been the culmination of 2 years of work towards making at least part of the quantum technology space truly open.

\subsection{Brief Overview of NVC Magnetometry}\label{sec:physics}

The best surveys of the physics behind Nitrogen-Vacancy (NV) centre diamond magnetometry are \cite{Rondin2014} for detailed theory, and \cite{Frellsen2018} for a high-level overview. 

\begin{figure}[t]
    \centering
    \includegraphics[width=0.7\linewidth]{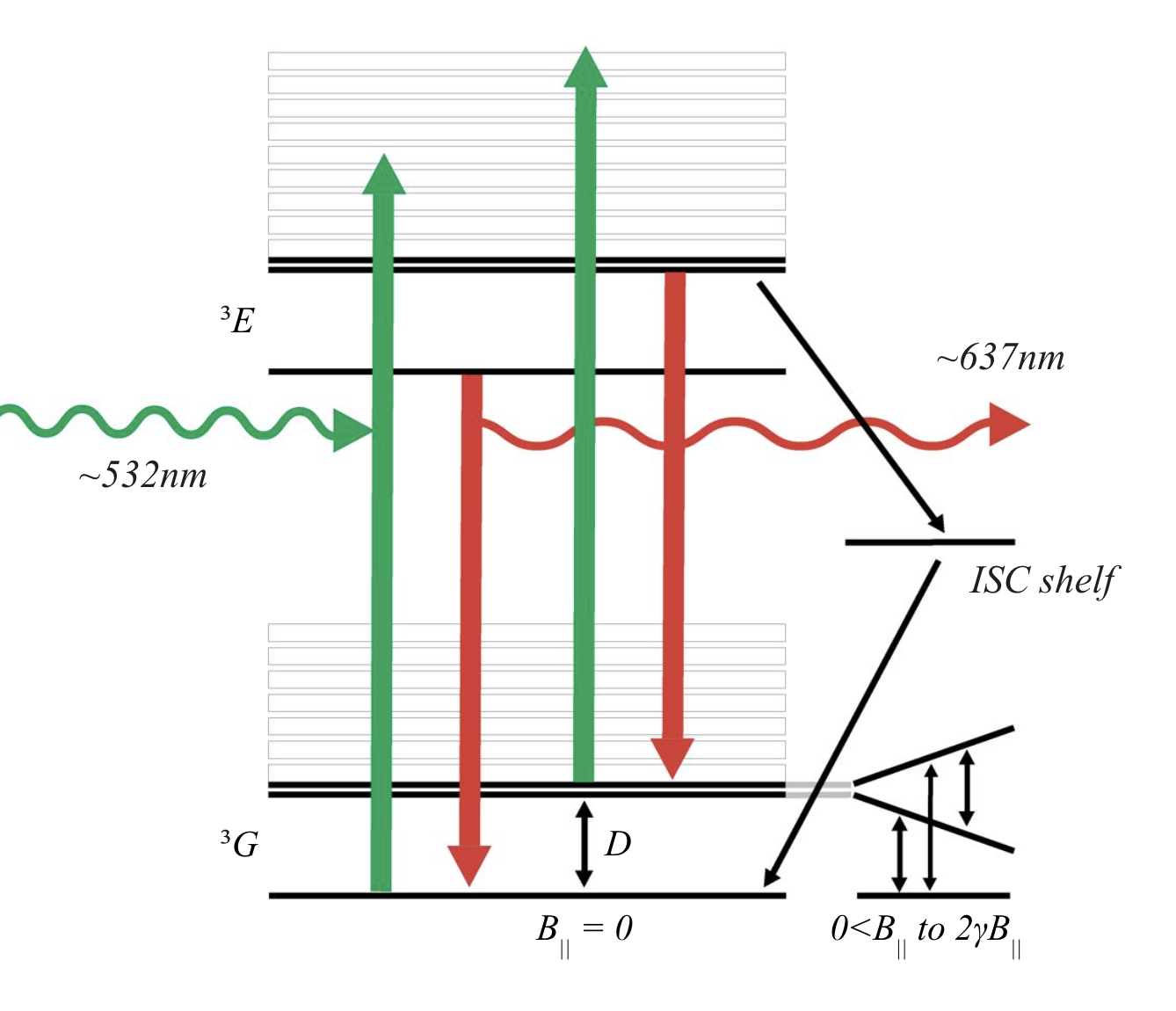}
    \caption{Energy Level diagram for NV Centre Magnetometry, detailed in section \ref{sec:physics}. {\bf NB} - the intersystem crossing (ISC) shelf involving two singlet states with an infrared (IR) transition occurring at 1042 nm, but this is not relevant to our magnetometry.}
    \label{fig:energy-levels}
\end{figure}

A Nitrogen Vacancy Centre (NV Centre) is the replacement in a diamond lattice of a single carbon (C) atom with a single Nitrogen (N) atom. In lab-grown diamonds this replacement is typically done through electron bombardment \cite{McLellan2016}. The mismatch in atomic valencies (4 for C to 3 for N) means that a vacancy is introduced. This defect essentially acts as a virtual atom contained within the lattice structure that has a strong zero-phonon line at 1.945 eV ($\lambda_{ZPL} = 637$nm) between the excited states $^3E$ and ground states $^3G$. The hyperfine structures at both these levels yield sub-levels that are split into a lower singlet state $m_s = 0$ and two upper states $m_s = \pm 1$ separated by a difference of $D=2.87$GHz with no magnetic field present \cite{Rondin2014}. The vacancy also has no Doppler shift relative to any interacting photons, which is why its effects may be utilized at room temperature.

The ground states Hamiltonian of this, ignoring the hyperfine interactions, is given in \cite{Stegemann2023} and \cite{Rondin2014} as:

\begin{equation}\label{hamiltonian}
\mathcal{H} = hDS^2_z + hE(S^2_y - S^2_x) + g \mu_B \mathbf{B} \cdot \mathbf{S} 
\end{equation}
where $z$ is the parallel axis to the NV centre, $D$ and $E$ are the Zero-Field Splitting parameters, $S^2_x$, $S^2_y$, and $S^2_z$ are the spin operators, $g$ is the Land\'{e} $g$-factor, and $\mu_B$ is the Bohr magneton. $D$ arises via spin-spin interaction between the two uncoupled electrons in the defect, and $E$ is proportional to the internal strain within the lattice. The resulting resonant frequency $\nu$ (see \cite{Stegemann2023}) for the applied microwave field at frequency $D$ is given by
\begin{equation}
    \nu = D \pm \sqrt{E^2 + (\gamma \cdot \mathbf{B}_{||})^2}
\end{equation}
with $\gamma = g\mu_B/h \approx 28$GHzT$^{-1}$ \cite{Rondin2014}. The important takeaway here is that under this simplified regime, provided $E$ is much less than $\mathbf{B}_{||}$, the main variable regarding the splitting is the magnetic field $\mathbf{B}_{||}$ and its orientation to the NV centre axis.

The Energy-level scheme found in figure \ref{fig:energy-levels} summarizes the key physical concepts. We can excite these NV centres with green light $\lambda_{E} \approx 532$nm. The NV centres then relax into the upper excited states $^3E$, and will become magnetosensitive under the application of microwaves. In practice, we sweep the microwave spectrum surrounding 2.87GHz, utilizing a similar sweep to the sweep in \cite{Stegemann2023} by starting at 2.614GHz and sweeping up to 3.126GHz in discrete steps of 4MHz. 

\section{Gap Analysis}

Many possible design options were reviewed, including the designs and build characteristics from \cite{Chatzidrosos2021,Dix2024,Homrighausen2023,Kuwahata2020,Opaluch2021,Stegemann2023,Webb2020}. The most complete guide was from Stegemann {\it et al.} \cite{Stegemann2023}, which provided a basis for our final design. 

Upon analysis, however, we found a number of key things were missing from the available designs; The firmware used in NV Center diamond experiments is generally not available. Likewise, a reliable photodetector design schematic is generally not available, with most examples in the literature not always accurate to best electronic design principles. There are few easily configurable diamond mounts, as most require specialist optical mounting apparatus. Lastly, none of the solutions were designed to be `hackable' for rapid deployment or prototyping, with little to no reconfigurable electronics and software. Similarly, none of the designs were geared to being ruggedized towards real world deployment.

\section{Design Overview}

In order to remediate these gaps, we came up with an initial design that made the following key decisions: 
\begin{enumerate}
    \item {\bf Use off the shelf components} - the only specialist part should be the diamond material and any custom PCBs.
    \item {\bf Make as many things `software problems' as possible} - by careful hardware design we can relinquish most control to a single software source. This massively improves debugging, maintenance, and updating a device.
    \item {\bf Simplify the design as much as possible.}
    \item {\bf Make the device `hackable'} - by this we mean that the device should be fundamentally and easily reconfigurable and/or extendable in both software and hardware. 
\end{enumerate}

\subsection{Microcontroller \& Firmware}

We wrote the source code for this device as Arduino IDE code - this taps into a large user base, with lots of help and built-in support for 26 native Arduino boards and hundreds of derivative boards \cite{Banzi2009-jg}. This does two things; first, it immediately makes the design much more accessible to more people who are familiar with Arduino, over using an IDE or development tool-chain that is manufacturer specific. 

Second, it allows the design to be ported to other chips through Arduino's universal hardware abstraction layer. This future-proofs the design, as any issues affecting the availability of one microcontroller means another more available one may be substituted with ease.

To this end, we used an ESP32 microcontroller - this is very well supported by Arduino IDE, has high quality 12-bit ADC inputs, is cheap and widely available, and has built-in WiFi and Bluetooth Low Energy (BLE) for future connectivity. 

The default software libraries for driving an ADF4351 chip (see below) use a short calculation loop to generate the input values for the chip based off the input of a desired frequency. However, the sensitivity of the electronics (see below) means that this interferes with measurements significantly. To mitigate this, we pre-computed all of the desired sweep values and store them in firmware. This only consumes around 30 bytes per frequency at most, and practically eliminates the noise from the microcontroller interfering with the measurement operations. 

\subsection{Electronics}

We made use of off-the-shelf operational amplifiers that are in a standard transimpedence configuration. We use the BPW34 photodiode as the main detection part as it has a large detection area - $3 \times 3$mm - and is widely available. This is then amplified by a TL082 dual op-amp, a common low-noise dual operational amplifier. 

To minimize the parts required and to only require one 5V power input, we utilize a voltage divider formed from two 1,000 Ohm resistors in series from 5V to Ground. This gives us a floating 2.5V which we utilize as a `fake ground' to which the non-inverting inputs (`$+$' pins) of the operational amplifiers are tied. This achieves two things - firstly, it means we can do away with additional inverting voltage regulators, and secondly means this guarantees that that output will be below the 3V limit on the ESP32's ADC input, without needing level shifting.

An ADF4351 PLL signal generator was chosen as a microwave generator as it allows control via SPI bus with a range of 35MHz to 4.4GHz. The output of this is then amplified by two monolithic microwave RF amplifiers, specifically GALI-84+ (or compatible) amplifiers providing $\approx 16$dB amplification each.

The PCBs were designed to use single-sided soldering and assembly of surface mount components, reducing cost and complexity of production. The PCBs also use top and bottom layer ground planes to reduce parasitics and noise.

Additionally, to make the device hackable beyond just giving full software access, we also incorporated a small prototyping area. These generally consist of 1.0mm copper (Cu) plated holes, spaced 2.54mm apart in a grid array. This allows the incorporation of any device which has output pins with this standard pitch, which is historically derived from the DIN 41612 standard, see \cite{Fletcher2013-ye}.

\subsection{Light Source}

Canonically, 525-532nm light is sourced from a laboratory laser, however these may sometimes prove to be problematic, especially where there is little to no guarantee of adequate laser safety protocols.

One of the main advantages of lasers is the generation of phase coherent photons. This is something we do not utilize for our  magnetometry (see equation \ref{hamiltonian}), so it suffices to use any source with the correct frequency output. `Superbrite' LEDs were sourced that emit the correct frequency, and were found to perform on par with laser sources. This significantly reduces cost and power consumption, whilst also increasing the safety of the device.

\subsection{Diamond and Photodiode Encapsulation}

\begin{figure}[t]
    \centering
    \includegraphics[width=0.74\linewidth]{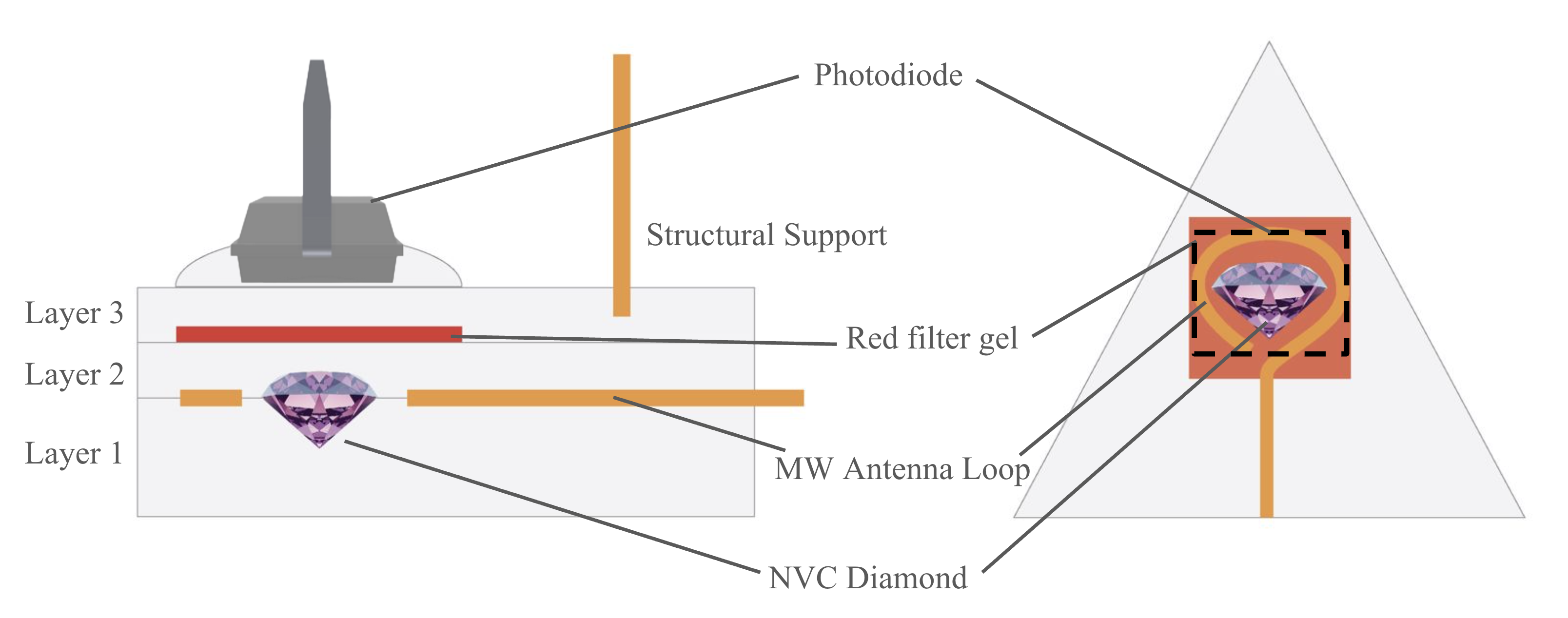}
    \includegraphics[width=0.25\linewidth]{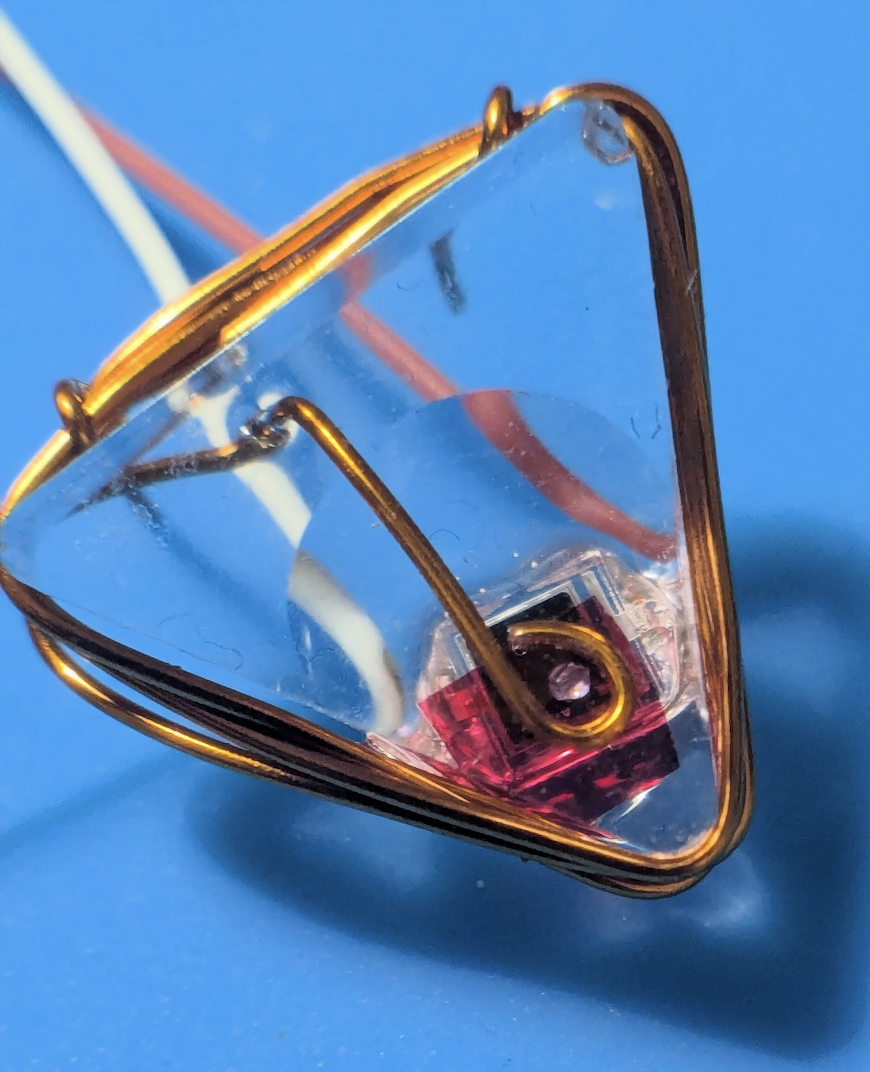}
    \caption{{\bf Left:} A diagram depicting the constructions layers of the diamon epoxy prism mount. {\bf Right:} A mounted sample of diamond.}
    \label{fig:mounted-diamond}
\end{figure}

We make use of fast-UV curing epoxy to encase the photodetector, a microwave aerial, and NV centre diamond samples. As pointed out in \cite{Stegemann2023}, we can omit the need for a dichroic mirror and instead place a square of easily available red lighting filter gel between the photodiode and diamond to act as our primary optical filter. 

The epoxy prism is created during a 3-step layering process, corresponding to the diagram in figure \ref{fig:mounted-diamond}; Layer 1 is a base layer of epoxy upon which the diamond and microwave aerial are placed after it is cured. A second layer of epoxy is then applied to fix the diamond and aerial in place. A third layer that fixes in place the filter, photodiode, as well as any mechanical fixing parts. 

The density of diamond means that it will naturally sink in most epoxies, hence a layering approach is used. Even for fast UV curing epoxy, the excellent optical properties of clear epoxy are not compromised by the layering technique. It is also easy to reconfigure, as most epoxies only need to be heated to $\sim200$C in order to delaminate. The diamond is completely unaffeced by this heating, and so can be retrieved for subsequent encapsulation. In particular, this move away from the optics table towards very low cost epoxy mounting is a key part of the fast iteration and innovation potential for this design.

\begin{figure}[t]
    \centering
    \includegraphics[width=0.5\linewidth]{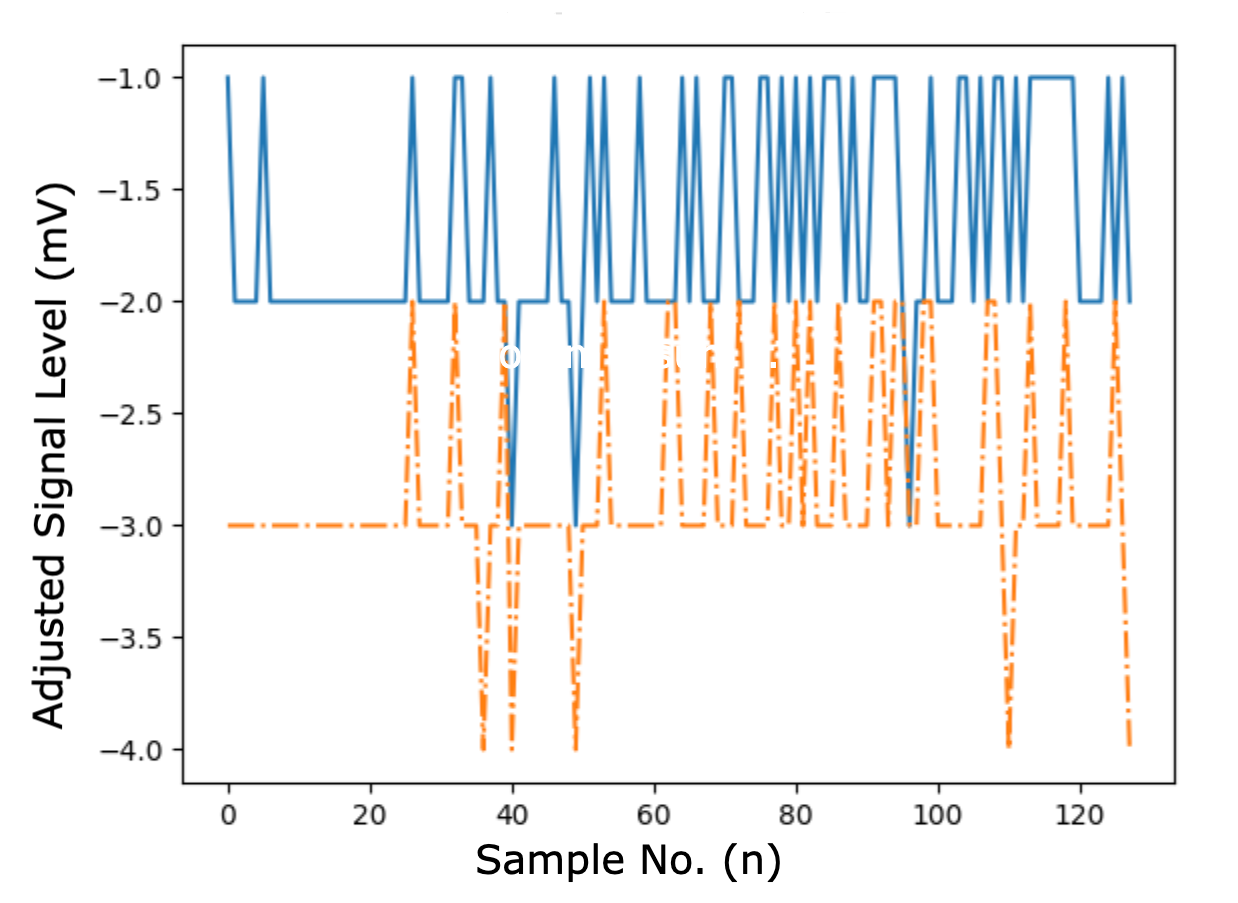}
    \includegraphics[width=0.3\linewidth]{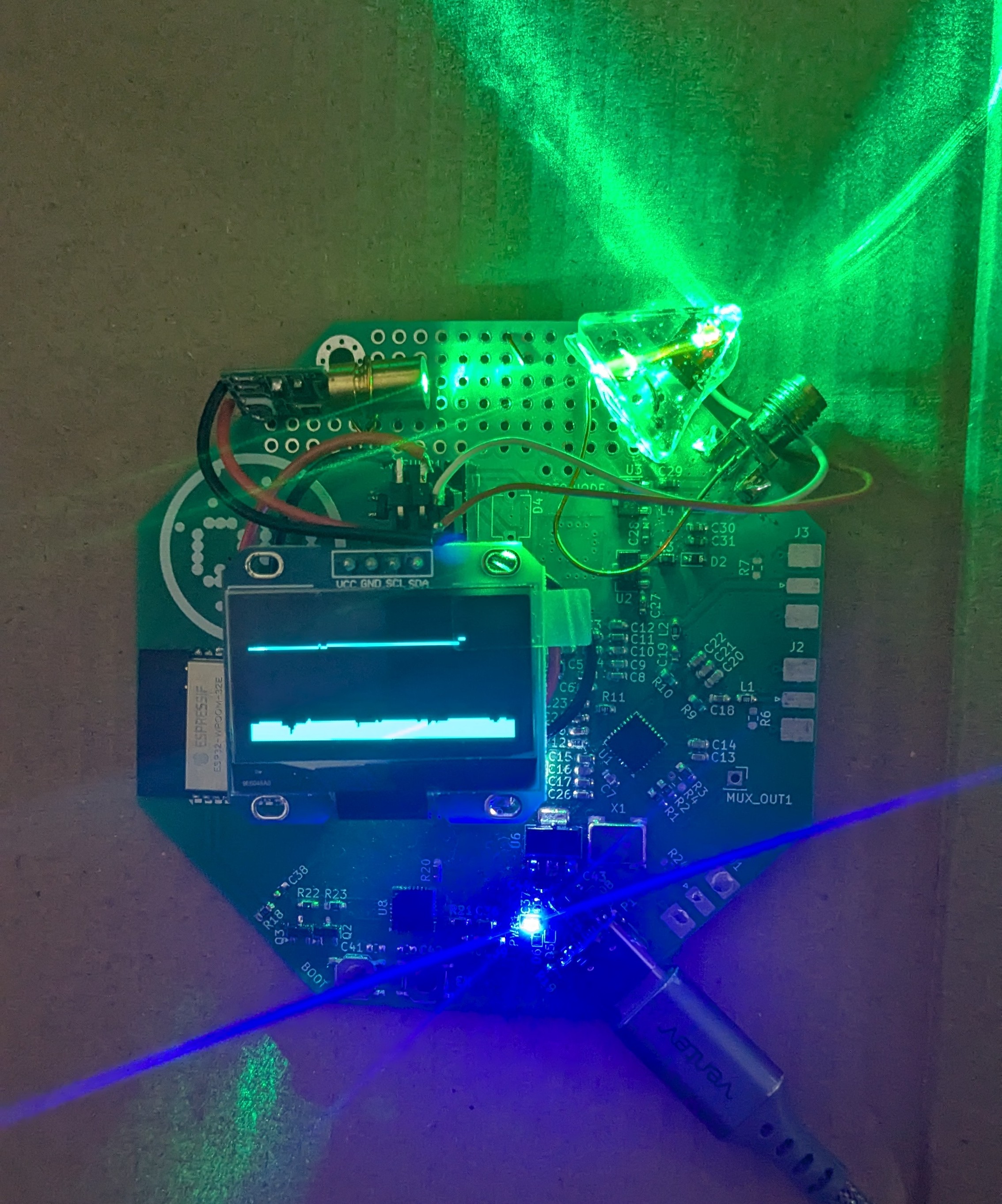}
    \includegraphics[width=0.5\linewidth]{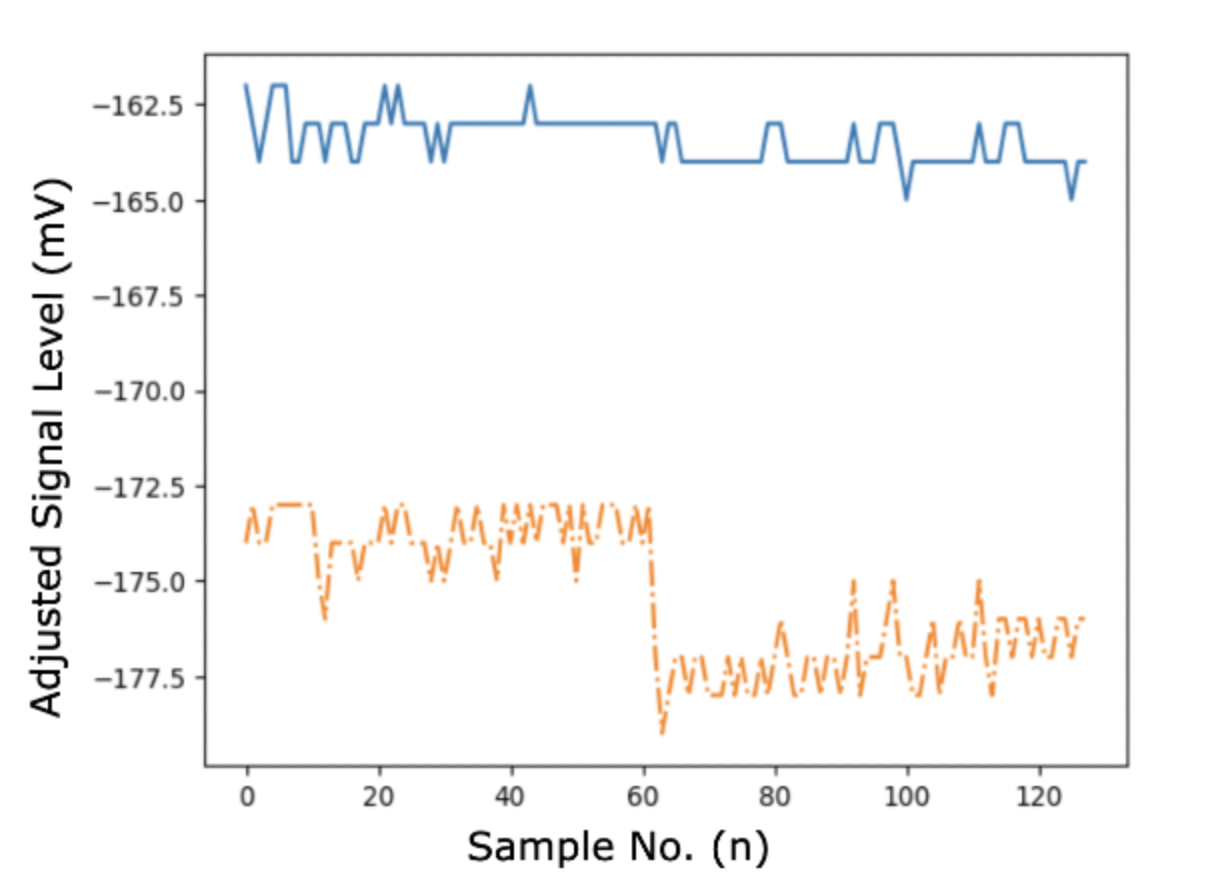}
    \caption{{\bf Upper Left:} Plotted readings in noise-adjusted millivolts from the ESP32 via serial over USB. Each point is an average of 6 measurements, with the dashed orange showing measurements with, and blue lines measurements without the presence of a small magnet. The shift in the splitting can be observed both in the drop in levels and in the shift of the valleys moving further apart. The frequency for each measurement is given in MHz from the sample number $n$ by $(2838 +n)$. {\bf Upper Right:} The device during operation. {\bf Lower:} Plotted readings in noise-adjusted millivolts with the sensor in the presence of a large magnet (dashed orange) compared to no magnet (blue). Same scale as other plot, but in a different noise environment, hence lower adjusted readings.}
    \label{fig:arduino-chart}
\end{figure}

\section{Testing}

Although full characterization of the quantum sensor is still yet to be performed, we have managed to get some preliminary checks and estimates of sensitivity. In figure \ref{fig:arduino-chart} we can see the output from the sensor both with and without the presence of a small magnet placed on top of the diamond prism. These are in line with the measurement of the magnetism of a small bar magnet, $\approx 0.005-0.01$ Teslas, found in figure 7 in \cite{Stegemann2023} and figure 3 in \cite{Rondin2014}, by rough approximation.


\section{Conclusions and Future Work}

\begin{table}[t]
\centering
\begin{tabular}{
>{\columncolor[HTML]{FFFFFF}}l 
>{\columncolor[HTML]{FFFFFF}}c 
>{\columncolor[HTML]{FFFFFF}}r }
\hline
\textbf{Item}          & \textbf{Number} & \textbf{Cost} \\ \hline
Battery pack           & 1               & Free to £5    \\ \hline
OpAmp + electronics    & 1               & up to £5      \\ \hline
LED                    & 1               & $\sim$£2-£5   \\ \hline
ADF4350/1 MW Generator & 1               & £15-25        \\ \hline
MW Gain (40dB) + BNC   & 1               & £10           \\ \hline
PCB (unit from run)    & 1               & $\sim$£5      \\ \hline
OLED Screen            & 1               & £5            \\ \hline
ESP32 board            & 1               & $\sim$£10     \\ \hline
NV Center Diamond      & 1               & $\sim$£10     \\ \hline
Epoxy and Cu Wire      & 1 ea.           & £10-15        \\ \hline
                       & \textbf{TOTAL}  & $\sim$£115    \\ \cline{2-3} 
\end{tabular}
\caption{Bill of Materials for the first release version sensor.}
\label{tab:bom}
\end{table}

On April 14th 2025 we released the full stack as open source on GitHub in \cite{UncutGem2025}. To this end, the schematics, PCBs, parts lists, software code, and build instructions are all fully available, free and open source. We hope to build on this success with future releases and updates to both the hardware and base software to enable more people to learn about and practice with quantum sensing. Table \ref{tab:bom} details the Bill of Materials (BoM) for the first release of the device, detailing a total cost that should not exceed £120. In Aug 2025 we released a single-board variant at DEF CON 33 \cite{MCVKAug2025,WIREDAug25} with a lower total price of just under \$100. With our subsequent version releases we expect this cost to fall even further. 

\subsection{Future Work}


Firstly, a full and accurate characterization should be performed on multiple diamond samples in order to gather data for both assessing the direct measurement and further analysis. There is plenty of scope for improving the output analysis and measurement estimation using edge machine learning and TinyML, something that Homrighausen {\it et al.} demonstrated in \cite{Homrighausen2023}, and which we think would be an important step forward in improving the output of these devices.

We would like to see a space-ready version of the device prepared. This would mean changing the form factor to that of a CubeSat, as well as ensuring that all of the materials, soldering, conformal coating, and mechanical properties are suitable for the stresses and strains of space. 

There are many use cases that should also be further explored for their suitability to this kind of low-cost quantum sensing setup, including; medical technology use cases like magnetocardiography (MCG), industrial sensing applications, and low-cost magnetometry for Position, Navication and Timing (PNT) use cases, building on work such as Muradoglu {\it et al.} \cite{Muradoglu2025}.

\begin{credits}
\subsubsection{\ackname} The authors are grateful to the following people for their help during development of the device: Prof. Ben Varcoe, Max Shirokawa-Aalto, David Benoit, Matthew Markham, Rick Altherr, David Bengtson, and Brian McDermott.

\subsubsection{\discintname}
The authors have no competing interests to declare that are relevant to the content of this article.
\end{credits}

\bibliographystyle{splncs04} 
\bibliography{refs}

\end{document}